\documentclass{JHEP3}
\usepackage{graphics}

\def\p{\partial}

\def\half{{1\over 2}}
\def\tilde{\widetilde}

\def\ah{{\widehat\alpha}}

\def\bh{{\widehat\beta}}
\def\gh{{\widehat\gamma}}

\def\Str{{\textrm{Str}}}
\def\a{\alpha}
\def\b{\beta}
\def\g{\gamma}
\def\d{\delta}
\def\e{\epsilon}

\def\vp{\varphi}

\def\l{\lambda}

\def\p{\partial}
\def\t{{\theta}}

\def\ah{{\hat \alpha}}

\def\ad{{\dot \a}}

\def\nn{{\nonumber}}
\def\ZZ{\ensuremath{\mathbb{Z}}}
\def\RR{\ensuremath{\mathbb{R}}}

\def\M{{\cal M}}

\newcommand{\QQ}{{\mbox{\small \bf Q}}}

\newcommand{\HH}{{\mbox{\small \bf H}}}

\newcommand{\EE}{{\mbox{\small \bf E}}}

\newcommand{\be}{\begin{equation}}
\newcommand{\ee}{\end{equation}}
\newcommand{\bea}{\begin{eqnarray}}
\newcommand{\eea}{\end{eqnarray}}

\title{The Pure Spinor Formulation of Superstrings}

\author{Yaron Oz\\
Raymond and Beverly Sackler School of Physics and Astronomy \\
Tel-Aviv University, Ramat-Aviv 69978, Israel\\}

\abstract{In this lectures we outline the construction of pure spinor
superstrings. We consider both the open and closed pure spinor superstrings in critical and noncritical
dimensions and on flat and
curved target spaces with RR flux.
We exhibit the integrability properties of pure spinor superstrings on curved
backgrounds with RR fluxes.\\

\vskip 1cm
{\it These lectures have been given in the
RTN Winter School on Strings, Supergravity and Gauge Theories, CERN (2008).}

}

\begin{document}

\newpage

\section{Introduction}

There are currently three main formulations of superstrings: The
Ramond-Neveu-Schwarz (RNS), the Green-Schwarz (GS) and the Pure Spinor (for a review see
\cite{Berkovits:2002zk}).
In the RNS formalism one studies maps $(x^m,\psi^m; m=0,...,9)$ from a two-dimensional supersymmetric
worldsheet to a bosonic space-time. The formalism lacks manifest space-time
supersymmetry and requires the introduction of a projection (GSO) in order to exhibit
it. In this formalism the $(2,2)$ worldsheet supersymmetry is related to space-time supersymmetry.

There are various  complications in the perturbative analysis in the  RNS formalism, such as a requirement
for summation
over spin structures and a lack of
a proper definition of the measure of integration on the supermoduli space.
The RNS formalism is also inadequate for the quantization on backgrounds with RR fluxes, i.e. there is no
simple coupling to the RR fields.

In the GS formalism we consider maps $(x^m , \theta^{\alpha}; m=0,...,9; \alpha=1,...16)$
 from a two-dimensional bosonic
worldsheet to a supersymmetric space-time.
This formalism posses a manifest space-time supersymmetry and can be used
to quantize superstrings on RR backgrounds.
It is, however, difficult to analyze the GS
quantum sigma-model.
The formalism
requires a gauge fixing of a fermionic symmetry ($\kappa$-symmetry),
which is known only in the light-cone gauge and hence
non-covariantly. Since the equations of motion of the GS superstring
do not provide a propagator for the $\theta$'s, the calculations in
worldsheet perturbation theory are problematic.

In the pure spinor formalism, as in the GS formalism,
we consider maps $(x^m , \theta^{\alpha}; m=0,...,9; \alpha=1,...16)$
 from a two-dimensional bosonic
worldsheet to a supersymemtric space-time.
We introduce additional new degrees of freedom, which are bosonic
target space  spinors (ghosts) $\lambda^{\alpha}; \alpha = 1,...16$. They satisfy
a set of constraints that define then as pure spinors, hence the name of the superstring.
The pure spinor sigma-model can be quantized in a straightforward manner,
since it contains additional terms that break explicitly the GS
$\kappa$-symmetry and introduce propagators for all the
variables.
This formalism posses a manifest space-time supersymmetry and can be used
to quantize superstrings on RR backgrounds.

In this lectures we will outline the pure spinor formulation of superstrings.
The lectures are organized as follows. In section 2 we will construct the pure spinor superstring in a flat ten-dimensional
space. We will consider both the open and closed pure spinor superstrings and the relation to the RNS superstring.
In section 3 we will consider the pure spinor superstring in curved space.
After a discussion of general curved backgrounds, we will consider Anti de-Sitter (AdS) backgrounds with RR flux.
In section 4 we will present noncritical pure spinor superstrings in various dimensions. We will discuss
various examples: the linear dilaton background, $AdS_2$ and $AdS_4$.
In section 5 we will exhibit the integrability properties of pure spinor superstrings on curved
backgrounds with RR fluxes.
In the appendix we give some details on superalgebras and supergroups.

\section{Pure Spinor Superstring in Flat Space}

In this section we will construct the pure spinor superstring in a flat ten-dimensional
target space \cite{Berkovits:2000fe}.

\subsection{Pure spinor  open superstring}

We will start with the construction of
the open pure spinor superstring, or more precisely the holomorphic part of the closed pure spinor superstring.

\subsubsection{The pure spinor superstring variables}

Consider the supermanifold
$(x^m,\theta^{\alpha})$, where $x^m, m=0,...,9$ are commuting
coordinates with the OPE
\be
x^m(z)x^n(0) \sim -\eta^{mn} log|z|^2 \ ,
\ee
and $\theta^{\alpha}, \alpha=1,...,16$ are
worldsheet weight zero anti-commuting coordinates.
$x^m$ transform in the vector representation of the target space Lorentz group $SO(1,9)$, while
$\theta^{\alpha}$ transform in its ${\bf 16}$ Majorana-Weyl spinor representation.

One introduces the worldsheet weight one $p_{\alpha}$ as the
conjugate momenta to $\theta^{\alpha}$ with the OPE \be
p_{\alpha}(z) \theta^{\beta}(0) \sim \frac{\delta_{\a}^{\b}}{z} \
. \ee
$(p_{\alpha}, \theta^{\beta})$ is a free fermionic $(b,c)$ system of weight $(1,0)$.
$(x^m, p_{\alpha}, \theta^{\beta})$ are the GS variables.

Next we add a bosonic complex Weyl spinor ghost $\lambda^{\alpha},\alpha=1,...,16$, which
satisfies the pure spinor constraint \be
\lambda^{\alpha}\gamma_{\alpha\beta}^m\lambda^{\beta} = 0 \;\;\;\;\;\; m=0,...,9 \ .
\label{const} \ee
The $\gamma_{\alpha\beta}^m$ are the symmetric
$16\times 16$ Pauli matrices in ten dimensions.
A spinor $\lambda^{\alpha}$ that satisfies the constraints (\ref{const}) is called pure spinor.
This set of ten constraints is reducible as we will soon discuss. It reduces the number of degrees
of freedom of $\lambda^{\alpha}$ from sixteen to eleven.

Another definition of the pure
spinors in even dimension
$d=2n$, which is due to Cartan and Chevalley is
\be
\l^{\a}\gamma^{m_1\ldots m_j}_{\a\b}\l^{\b} = 0 \;\;\;\;\;\;\;\;  j<n  \ ,
\label{CC}
\ee
so that
the pure spinor bilinear reads
 \bea
 \l^\a\l^\b={1\over n!2^n} \gamma^{ \a\b}_{m_1\ldots
 m_n}(\l^{\sigma}\gamma^{m_1\ldots m_n}_{\sigma\tau}\l^{\tau}) \ , \label{cartan}
 \eea
where $\gamma^{m_1\ldots m_j}$ is the antisymmetrized product of
$j$ Pauli matrices. The pure spinor that we consider in ten dimensions satisfies (\ref{cartan})
with $n=5$.

We note, for later use when we will discuss pure spinor superstrings in various dimensions,  that this definition of the pure spinor space in
$d=2,4,6$ dimensions is trivially realized by an $SO(d)$ Weyl
spinor.

We denote the conjugate momenta to $\lambda^{\alpha}$ by the worldsheet weight one bosonic target space complex Weyl spinor $ w_\a$.
The system $(w_\a,\lambda^{\alpha})$ is a curved $(\beta,\gamma)$ system
of weights $(1,0)$. The reason that the system is not free is the set of pure spinor
constraints (\ref{const}).

The pure spinor constraints imply that $ w_\a$ are defined up to
the gauge transformation \be \delta  w_\a = \Lambda^m (\gamma_m
\lambda)_{\alpha} \ . \ee Therefore, $ w_\a$ appears only in gauge
invariant combinations. These are the Lorentz algebra currents
$M_{mn}$, the ghost number current $J_{(w,\lambda)}$ which assigns
ghost number $1$ to $\lambda$ and ghost number $-1$ to $ w $ \be
M_{mn} = \frac{1}{2}w\gamma_{mn}\lambda,~~~~~J_{(w,\lambda)} =
 w_\a\lambda^{\alpha} \ , 
\label{lorentz} 
 \ee and the pure spinor
stress-energy tensor $T_{(w,\lambda)}$.

Unlike the RNS
superstrings, all the variables that we use in the pure spinor
superstring are of integer worldsheet spin.
This is an important property of the formalism: for instance,  we will
have no need to
sum over spin structures when computing multiloop scattering amplitudes.

\subsubsection{The pure spinor space}

The pure spinor
set of constraints (\ref{const}) defines a curved space, which can be covered by sixteen patches
$U_{\alpha}$ on which the $\a th$ component of $\l^{\a}$ is nonvanishing.

The set of constraints   (\ref{const}) is reducible.
In order to solve it we rotate to Euclidean signature. The pure spinor
variables $\lambda^\a$ transform in the ${\bf 16}$ of $SO(10)$.
Under $SO(10) \rightarrow U(5) \simeq SU(5)\times U(1)$ we have
that ${\bf 16} \rightarrow {\bf 1}_{\frac{5}{2}} \oplus {\bf \bar{10}}_{\frac{1}{2}} \oplus {\bf 5}_{-\frac{3}{2}}$.
We denote the sixteen components of the pure spinor in the
$U(5)$ variables by $\lambda^{\alpha} = \lambda^{+}\oplus\lambda_{ab}\oplus \lambda^a ; a,b=1..5$ with
 $\lambda_{ab}= -\lambda_{ba}$.
In this variables it is easy to solve the pure spinor
set of constraints (\ref{const}) by
\be
\lambda^{+} = e^s,\;\;\;\; \lambda_{ab}= u_{ab},\;\;\;\; \lambda^a = -\frac{1}{8}e^{-s} \varepsilon^{abcde}u_{bc}u_{de} \ .
\label{U5}
\ee

The pure spinor space  ${\cal M}$
is complex eleven-dimensional,which is a cone over
${\cal Q} = \frac{SO(10)}{U(5)}$.
At the origin $\lambda^\a = 0$, both the  pure spinor
set of constraints (\ref{const}) and their derivatives with respect to $\lambda^\a$ vanish.
Thus, the pure spinor space has a singularity at the origin.

\subsubsection{The pure spinor superstring action}

We will work in the worldsheet conformal gauge.
The conformal gauge fixed worldsheet action of the pure spinor superstring is $S=S_0+S_1$, where \be S_0 =
\int d^2z \left(\frac{1}{2}\p x^m\bar{\p}x_m +
p_{\a}\bar{\p}\theta^{\alpha} -   w_\a\bar{\p} \lambda^{\alpha}
\right) \ , \ee and \be \label{topcop} S_1 = \int d^2z
\left(\frac{1}{4}r^{(2)}\log \Omega(\lambda)\right) \ . \ee
The first two terms in $S_0$ correspond to the GS action written in a first order formalism,
the third term in $S_0$ is the pure spinor action, and
$S_1$
is a coupling of the worldsheet curvature $r^{(2)}$ to the
holomorphic top form $\Omega$ of the pure spinor space ${\cal M}$
 \be \Omega = \Omega(\lambda)d \l^1\wedge ...\wedge
d\l^{11} \ .
 \ee
 Note that the
$(w_\a,\lambda^{\alpha})$ action is holomorphic and does not
depend on their complex conjugates.

The stress tensor of the $(w,\lambda)$ system reads
 \be
T_{(w,\lambda)} =   w_\a\partial \lambda^{\alpha} -
\frac{1}{2}\p^2\log \Omega(\l) \ , \label{Ttop}
 \ee
and we will discuss the significance of the last term later. Note, however, that
it does not contribute to the central charge of the system.

The
system $( w_\a,\lambda^{\alpha})$ is interacting due to the pure
spinor constraints.
It has the central charge $c_{( w ,\lambda)} =
22$, which is twice the complex dimension of the pure spinor space
\be T_{( w ,\lambda)}(z)T_{( w ,\lambda)}(0) \sim
\frac{dim_{\mathbb{C}}({\cal M})}{z^4} + ... \ . \ee
This can be computed, for instance,  by introducing
the conjugate momenta to the $U(5)$ variables (\ref{U5}) with the OPE
\be
t(z)s(0) \sim log z,\;\;\;\;\;\;\;\; v^{ab}u_{cd} \sim -\frac{\delta^{ab}_{cd}}{z} \ ,
\ee
with
$\delta^{ab}_{cd} = \frac{1}{2}(\delta^a_c \delta^b_d - \delta^a_d \delta^b_c)$.
The stress energy tensor reads
\be
T_{( w ,\lambda)} = v^{ab}\p u_{ab} + \p t \p s + \p^2 s \ ,
\ee
giving the central charge $22$.

The total central charge of the pure spinor superstring is
\be
c^{tot} = c(X^m) + c(p_{\a},\theta^{\alpha}) + c(w_\a,\lambda^{\alpha})  = 10-32+22=0 \ ,
\ee
as required by the absence of a conformal anomaly.

The ghost number anomaly reads \be J_{( w ,\lambda)}(z)T_{( w
,\lambda)}(0) \sim -\frac{8}{z^3} + ... = \frac{c_1({\cal
Q})}{z^3}+ ...\ , \ee where ${c_1({\cal Q})}$ is the first Chern
class of the pure spinor cone base ${\cal Q}$.

\subsubsection{The BRST operator}

The physical states are defined as the ghost number one cohomology
of the nilpotent BRST operator \be Q = \oint dz\,
\lambda^{\alpha}d_{\a} \ , \label{10ld}\ee where \be d_{\a} =
p_{\a} - \frac{1}{2}\gamma^m_{\a\b}\theta^{\b}\p x_m
-\frac{1}{8}\gamma^m_{\a\b}\gamma_{m\g\delta}\theta^{\b}
\theta^{\g}\p\theta^{\delta} \ . \ee This BRST operator is an
essential ingredient of the formalism, however it is not clear how to
derive its form by a gauge fixing procedure.
$Q^2=0$ since  $d_{\a} d_{\b} \sim \gamma^m_{\a\b}$ and $\l^{\a}\gamma^m_{\a\b}\l^{\b}=0$.

The $d_{\a}$ are the supersymmetric Green-Schwarz constraints.
They are holomorphic and satisfy the OPE
 \be d_{\a}(z) d_{\b}(0)
\sim - \frac{\gamma^m_{\a\b}\Pi_m(0)}{z} \ , \ee and \be d_{\a}(z)
\Pi^m (0) \sim \frac{\gamma^m_{\a\b}\p \theta^{\b}(0)}{z} \ , \ee
where \be \Pi_m = \p x_m + \frac{1}{2} \theta\gamma_m \p \theta \
, \ee is the supersymmetric momentum. $d_{\a}$ acts on function on
superspace $F(x^m, \theta^{\a})$ as \be d_{\a}(z)
F(x^m(0),\theta^{\b}(0)) \sim \frac{D_{\a}
F(x^m(0),\theta^{\b}(0))}{z} \ , \ee where \be D_{\a} =
\frac{\p}{\p \theta^{\a}} +
\frac{1}{2}\gamma^m_{\a\b}\theta^{\b}\p_m \ , \ee is the
supersymmetric derivative in ten dimensions.

\subsubsection{Massless States}

Massless states are described by the ghost number one weight zero
vertex operators \be {\cal V}^{(1)} =
\lambda^{\alpha}A_{\alpha}(x,\theta) \ , \ee
where $A_{\alpha}(x,\theta)$ is an unconstrained spinor superfield.

The BRST cohomology conditions are
\be
Q{\cal V}^{(1)}=0, {\cal V}^{(1)} \simeq {\cal V}^{(1)} +
Q \Omega^{(0)} \ ,
\label{coho}
\ee
where $\Omega^{(0)}$ is a real scalar superfield.
These imply the ten-dimensional field equations for $A_{\alpha}(x,\theta)$
\be
\gamma^{\a\b}_{mnpqr}D_{\a}A_{\beta}(x,\theta) = 0 \ ,
\ee
with the gauge transformation $\delta A_{\a} = D_{\a}\Omega^{(0)}$, and we used the relation
(\ref{cartan}) in ten dimensions.

These equations imply that $A_{\alpha}$ is
an on-shell super Maxwell spinor superfield in ten dimensions
\be
A_\alpha (x, \theta) = \frac{1}{2} (\gamma^m \theta)_\alpha a_m (x) +
\frac{i}{12} (\theta \gamma^{mnp} \theta) (\gamma_{mnp})_{\alpha
\beta} \psi^\beta (x) + O(\theta^3)
\ee
where $a_m(x)$ is the gauge field and $\psi^{\b}(x)$ is the
gaugino. They satisfy the super Maxwell equations
\be
\partial^m(\p_m a_n -\p_n a_m) = 0,\;\;\;\;\;\;  \gamma_{\a\b}^m \p_m \psi^{\beta} = 0 \ .
\ee

$A_{\alpha}$ is related to the gauge superfield $A_m$ by \be A_m =
\g_{m}^{\a\b}D_{\a}A_{\b} \ , \ee and $A_m(x,\theta) = a_m(x) +
O(\theta) $. Only in ten dimensions do these conditions give an
on-shell vector multiplet. In lower dimensions they describe an
off-shell vector multiplet.

The integrated ghost number zero vertex operator for the massless
states reads \be {\cal V} =  \int dz U = \int dz \left(\p\theta^{\a}A_{\a} +
\Pi^m A_{m} + d_{\a}W^{\a} + \frac{1}{2}M_{mn}F^{mn}\right) \ ,
\ee where $W^{\a}$ and $F^{mn}$ are the spinorial and bosonic
field strength, respectively
\be
F_{mn} = \p_m A_n -\p_n A_m,\;\;\;\;\;\; D_{\a}W^{\b} = \frac{1}{4}(\gamma^{mn})^{\b}_{\a} F_{mn} \ .
\ee
 We have $W^{\a} = \psi^{\a} + ...$ and $ F_{mn} = f_{mn} +... $.

$M_{mn}$ are the generators of Lorentz transformations (\ref{lorentz}).
When expanded $U$ reads
\be
U = a_m \p x^m + \frac{1}{2}f_{mn}M^{mn} + \psi^{\a}q_{\a} + ...
\ ,
\ee
where $q_{\a}$ is the space-time supersymmetry current
\be
q_{\a} = p_{\a} +  \frac{1}{2}(\p x^m + \frac{1}{12}\theta\gamma^m\p \theta)(\gamma_m\theta)_{\a} \ .
\ee

Note that, unlike the RNS,  we did not need a GSO projection in order to get a supersymmetric spectrum
and that the Ramond and NS sectors appear on equal footing in the pure spinor formalism.

\subsubsection{Massive States}

The analysis of the massive states
proceeds in a similar way. At the first massive level $(m^2\a' = 1)$, the ghost
number one weight one vertex operator has the expansion \cite{Berkovits:2002qx}
\be {\cal U}^{(1)} = \p\lambda^{\a}A_{\a}(x,\theta)
+   \lambda^{\a}\p \theta^{\b}B_{\a\b}(x,\theta) +  d_{\b}\l^{\a} C_{\a}^{\b}(x,\theta) + ... \ . \ee
It describes a massive spin two multiplet with 128 bosons and 128 fermions
with $g_{mn}$ tracless symmetric
\be
\eta^{mn}g_{mn} =0,\;\;\;\;\;\; \p^m g_{mn}=0 \ ,
\ee
$b_{mnp}$ a 3-form
\be
\p^m b_{mnp} = 0 \ ,
\ee
and spin $3/2$ field $\psi_{m\a}$
\be
\p^m \psi_{m\a} = 0,\;\;\;\;\;\; \gamma^{m\a\b}\psi_{m\b} =0 \ .
\ee
This multiplet fields can be understood as the Kaluza-Klein modes of the eleven-dimensional
supergravity multiplet.

The procedure
at the nth massive level $m^2 = \frac{n}{\alpha'}$ is to construct a
ghost number one  weight $n$  vertex operator at zero momentum by using
the building blocks $(\p x^m,\theta^{\a},\l^{\a})$ and $(d_{\a},M_{mn},J_{(w,\l})$
and impose the conditions (\ref{coho}).

\subsubsection{Scattering Amplitudes}

Consider the tree level open string scattering amplitudes.
The n-point function $A_n$ reads
\be
A_n = \langle V_1(z_1)V_2(z_2)V_3(z_3)\int dz_4 U(z_4)...\int dz_n U(z_n) \rangle \ .
\ee
$V_i$ are dimension one, ghost number one vertex operators and $U_i$ are
dimension zero, ghost number zero vertex operators.
Their construction has been discussed above.

We can use the $SL(2,R)$ symmetry to fix three of the worldsheet coordinates, and using the free field OPE's
obtain
\be
A_n = \int dz_4 ...\int dz_n \langle \l^{\a}\l^{\b}\l^{\gamma} f_{\a\b\gamma}(z_r,k_r,\theta) \rangle \ ,
\ee
where $k_r$ are the scattering momenta and we integrated over the nonzero modes.
Thus, $f_{\a\b\gamma}$ depends only on
the zero modes of $\theta$.
There were eleven bosonic zero modes of $\lambda^{\a}$ and sixteen fermionic zero modes
of $\theta^{\a}$.
One expects eleven of the integration fermionic zero modes to cancel  the eleven integration
bosonic zero modes, leaving five fermionic $\theta$ zero modes.
A Lorentz invariant prescription for integrating over the remaining five fermionic $\theta$ zero modes is given by \cite{Berkovits:2000fe}
\be
A_n = T_{\rho\sigma\tau}^{\a\b\gamma}
(\frac{\p}{\p\theta}\gamma^{lmn}\frac{\p}{\p\theta})
(\gamma_l\frac{\p}{\p\theta})^\rho
(\gamma_m\frac{\p}{\p\theta})^\sigma
(\gamma_n\frac{\p}{\p\theta})^\tau
\int dz_4 ...\int dz_n f_{\a\b\gamma}(z_r,k_r,\theta) \ ,
\ee
where
$T_{\rho\sigma\tau}^{\a\b\gamma}$ is symmetrized with respect to the upper and lower indices, and
$T_{\a\b\gamma}^{\a\b\gamma}=1$, $T_{\rho\sigma\tau}^{\a\b\gamma}\gamma^m_{\a\b} = T_{\rho\sigma\tau}^{\a\b\gamma}\gamma_m^{\rho\sigma}=0$.
For examples of explicit calculations of scattering amplitudes see e.g. \cite{Policastro:2006vt}.

\subsection{Pure spinor closed superstring}

The construction of the closed superstrings is straightforward.
One introduces the right moving superspace variables
 $(\bar{p}_{{\alpha}}, \bar{\theta}^{{\alpha}})$, the pure spinor
 system $(\bar{ w }_{{\alpha}},\bar{\lambda}^{{\alpha}})$ and
the nilpotent BRST operator \be \bar{Q} = \oint
d\bar{z}\,\bar{\lambda}^{{\alpha}}\bar{d}_{{\a}} \ . \ee
The analysis of the spectrum proceeds by combining the left and
right sectors.
In our notation we will use the same spinorial indices for the right sectors.

Massless states are described by the ghost number $(1,1)$ vertex operator
\be
V = \lambda^{\a}\bar{\lambda}^{{\alpha}}A_{\a{\b}}(X,\theta, \bar{\theta}) \ .
\ee
$A_{\a\b}(X,\theta, \bar{\theta})$ is the on-shell supergravity multiplet is ten dimensions
\bea
A_{\a\b}(X,\theta, \bar{\theta}) = h_{mn}(\gamma^m\theta)_{\a}(\gamma^m\bar{\theta})_{\b}
+\psi_n^{\gamma}(\gamma^m\theta)_{\a}(\gamma_m\theta)_{\gamma}(\gamma^n\bar{\theta})_{\b} +...\nonumber\\
+
F^{\gamma\delta}(\gamma^m\theta)_{\a}(\gamma_m\theta)_{\gamma}(\gamma^n\bar{\theta})_{\b}(\gamma_n\bar{\theta})_{\delta} +...\ ,
\eea
$h_{mn}$ is the graviton, $\psi_n^{\gamma}$ is the gravitino and  $F^{\gamma\delta}$ the RR fields
\be
F^{\gamma\delta} = \oplus F^{k_1...k_n}(\gamma_{k_1...k_n})^{\gamma\delta} \ .
\ee
We see that all the different RNS sectors (NS,NS), (R,R), (R,NS) and (NS,R) are on equal footing in this representation.

The integrated ghost number zero
vertex operator for the massless states reads \be {\cal U} = \int
d^2z \left(\p\theta^{\a}A_{\a{\b}}
\bar{\p}\bar{\theta}^{\b} + \p\theta^{\a} A_{\a
m}\bar{\Pi}^m
 + ... \right) \ .
\ee

\subsection{Anomalies}

The pure spinor system $(\lambda^{\alpha},w_{\alpha})$ defines a
non-linear $\sigma$-model due to the curved nature of the pure
spinor space (\ref{const}). There are global obstructions to
define  the pure spinor system on the worldsheet and on target
space \cite{Witten:2005px,Nekrasov:2005wg}. They are associated
with the need for holomorphic transition functions relating
$(\lambda^{\alpha},w_{\alpha})$ on different patches of the pure
spinor space, which are compatible with their OPE. They are
reflected by quantum anomalies in the worldsheet and target space
(pure spinor space) diffeomorphisms. The conditions for the
vanishing of these anomalies are the vanishing of the integral
characteristic classes \label{anomalies} \be
\frac{1}{2}c_1(\Sigma)c_1(\M) = 0,~~~~~~\frac{1}{2}p_1(\M)=0 \ ,
\ee $c_1(\Sigma)$ is the first Chern class of the worldsheet
Riemann surface, $c_1(\M)$ is the first Chern class of the pure
spinor space $\M$, and $p_1$ is the first Pontryagin class of the
pure spinor space. The vanishing of $c_1(\M)$ is needed for the
definition of superstring perturbation theory and it implies the
existence of the nowhere vanishing holomorphic top form
$\Omega(\l)$ on the pure spinor space $\M$, that appears in the
stress tensor (\ref{Ttop}).

The pure spinor space (\ref{const}) has a singularity at
$\lambda^{\alpha} = 0$. Blowing up the singularity results in an
anomalous theory. However, simply removing the origin leaves a
non-anomalous theory. This means that one should consider the pure
spinor variables as twistor-like variables. Indeed this is a
natural intrepretation of the pure spinor variables considering
them from the twistor string point of view.

\subsection{Mapping RNS to pure spinors}
\label{10mapsection}

In this subsection we will construct a map from the RNS variables to
the pure spinor ones \cite{Adam:2007ws,Berkovits:2001us}. We will make use of a parameterization of
the pure spinor components that would make the $\beta
\gamma$-system structure of the pure spinor variables explicit. In
this way we will gain a new insight into the global definition of
the pure spinor space and the importance of its holomorphic top
form. The pure spinor stress tensor we will obtain by the map will
contains the contribution of the holomorphic top form on the pure
spinor space.

In the following we will consider the holomorphic sector.
The holomorphic supercharges in the
$-\frac{1}{2}$ picture of the RNS superstring are given by the
spin fields
\begin{equation}
  q_s = e^{-\phi / 2 + \sum_{I=1}^5 s_I H^I} \ ,
\end{equation}
where the $H^I$s are the bosons obtained from the bosonization of
the RNS worldsheet matter fermions and the $s_I$'s take the values
$\pm\frac{1}{2}$. These supercharges decompose into two Weyl
representations.

In order to proceed with the map, one must first solve the pure
spinor constraint (\ref{const}), going to one
patch of the pure s[pinor space. In each patch a different component of the
pure spinor is non-zero. The field redefinition we will use maps
the RNS description into one patch of the pure spinor manifold.
We will work on one of the patches which is
described by the $SU(5) \times U(1)$ decomposition of
the pure spinor $\l^\a=(\l^+,\l^a,\l_{ab})$, discussed previously.
 The component of the
pure spinor assumed to be non-zero is $\lambda^+$ corresponding to
the representation ${\bf 1}_\frac{5}{2}$ of this decomposition. In this
patch one can solve for the ${\bf 5}_{-\frac{3}{2}}$ components
$\lambda^a$ in terms of $\lambda^+$ and the components in the
${\bf \bar{10}}_\frac{1}{2}$ representation $\lambda_{ab}$.

On this patch the supercharge $q_+$ which is the singlet of
$SU(5)$ is raised to the $+\frac{1}{2}$ picture:
\begin{equation}
  q_+ = b \eta e^{3 \phi / 2 + i \sum_a H^a / 2} + i \sum_a \partial
  (x^a + i x^{a+5}) e^{\phi / 2 + i \sum_b H^b / 2 - i H^a} \ ,
\end{equation}
while the supercharges $q_a$, corresponding to the pure spinor
components $\lambda^a$ we solved for, remain in the $-\frac{1}{2}$
picture. Together they form a part of the original ten-dimensional
supersymmetry algebra.  One then defines the fermionic momenta
\begin{equation}
  p_+ = b \eta e^{3 \phi / 2 + i \sum H^a / 2} \ , \quad p_a = q_a
\end{equation}
and their conjugate coordinates $\theta^+$ and $\theta^a$. Note
that the OPE's of the fermionic momenta among themselves  are all
non-singular.

Introduce two new fields $\tilde
\phi$ and $\tilde \kappa$ using
\begin{equation}
\label{field10}
  \eta = e^{\tilde \phi + \tilde \kappa} p_+ \ , \qquad b = \quad e^{(\tilde \phi
  - \tilde \kappa) / 2} p_+ \ ,
\end{equation}
yielding
\begin{eqnarray}
  \tilde \phi & = & - {3i\over4} \sum_a H^a -  \kappa - {9\over4}
   \phi + \half \chi \ ,\\
  \tilde \kappa & = & {i\over4}\sum_a H^a - \kappa - {3\over4}
  \phi - \half\chi  \ ,
\end{eqnarray}
whose OPE's are
\begin{equation}
  \tilde \phi (z) \tilde \phi (0) \sim - \log z \ , \quad \tilde
  \kappa (z) \tilde \kappa (0) \sim \log z \ .
\end{equation}
The reason why we choose the particular field redefinition
(\ref{field10}) is that the
pure spinor formalism is equivalent to the RNS
formalism when we take into account all the different pictures at
the same time, which is achieved by working in the large Hilbert
space, that is including the zero modes of the ghost $\xi$. But
the usual cohomology of the RNS BRST charge $Q_{RNS}$ in the small
Hilbert space is equivalent to the cohomology of
$Q_{RNS}+\oint\eta$ in the large Hilbert space. With the
redefinition (\ref{field10}), we are then mapping the $\oint\eta$
term of this extended BRST charge directly to the part $\oint
\l^+d_+$ of the BRST operator (\ref{10ld}).

By substituting the map into the RNS energy-momentum tensor one
obtains
\begin{eqnarray}
  T & = & T_\mathrm{m} + T_\mathrm{gh} = - \frac{1}{2} \sum_m
  (\partial x^m)^2 - p_+ \partial \theta^+ - \sum_a p_a \theta^a -
  \nonumber\\
  && {} - \frac{1}{2} (\partial \tilde \phi)^2 + \frac{1}{2} (\partial
  \tilde \kappa)^2 + \partial^2 \tilde \phi + \partial^2 \tilde \kappa \ .
  \label{eq:critical-energy-momentum}
\end{eqnarray}
This can be verified to have a vanishing central charge
\begin{displaymath}
  c = (10)_x + (-12)_{p \theta} + (2)_{\tilde \phi \tilde \kappa} = 0 \ .
\end{displaymath}
The pure spinors are reconstructed by the ordinary bosonization of
a $\beta \gamma$-system \cite{Friedan:1985ge}
\begin{equation}
  \lambda^+ = e^{\tilde \phi + \tilde \kappa} \ , \quad
  w_+ = \partial \tilde \kappa e^{-\tilde \phi - \tilde \kappa} \ ,
\end{equation}
whose OPE is
\begin{equation}
  w_+ (z) \lambda^+ (0) \sim \frac{1}{z} \ .
\end{equation}
But the naive stress tensor one would expect for this  $\beta
\gamma$-system
\begin{displaymath}
  w_+ \partial \lambda^+ = - \frac{1}{2} (\partial \tilde \phi)^2 +
  \frac{1}{2} (\partial \tilde \kappa)^2 - \frac{1}{2} \partial^2 \tilde
  \phi - \frac{1}{2} \partial^2 \tilde \kappa \ ,
\end{displaymath}
does not coincide with the one we got from the map
(\ref{eq:critical-energy-momentum}). This shows that the pure
spinor stress tensor is not simply $w_+\p\l^+$ but actually
\begin{equation}
  T_\lambda = w_+ \partial \lambda^+ - \frac{1}{2} \partial^2 \log
  \Omega(\l) \ ,
\end{equation}
where $\Omega$ is the coefficient of a top form defined on the
pure spinor space \cite{Nekrasov:2005wg}. By comparison we can
read off the top form itself
\begin{equation}\label{topp}
  \Omega = e^{-3 (\tilde \phi + \tilde \kappa)} = (\lambda^+)^{-3} \ .
\end{equation}

At this point, we can map the RNS saturation rule for amplitudes
on the sphere
 \be
 \langle c\p c\p^2 c e^{-2\phi}\rangle=1,
 \ee
to the pure spinor variables, obtaining
 \be
 \langle (\l^+)^3(\t^a)^5\rangle=1,
 \ee
which is the prescription for the saturation of the zero modes in
that we used when discussing the scattering amplitudes. Note that the
third power of the pure spinor is consistent with the expression
of the holomorphic top form (\ref{topp}) we just reconstructed.

In the final step in performing the map we
covariantize by adding the missing coordinates
and momenta. We add a BRST
quartet consisting of ten $(1, 0)$ $b c$-systems $(p_{ab},
\theta^{ab})$ and ten $(1, 0)$ $\beta \gamma$-systems $(w_{ab},
\lambda^{ab})$. They have opposite central charges, so the total
central charge remains unchanged. In this way we recover the full
pure spinor stress tensor
 \be
 T=-\half\p x^m\p
 x_m-d_\a\p\t^\a+w_\a\p\l^\a-\half\p^2\log\Omega(\l).
 \ee

\section{Pure Spinor Superstring in Curved Space}

In this section we will consider critical pure spinor superstrings in curved ten-dimensional
spaces.

\subsection{General curved backgrounds}

The pure spinor action in curved target space is obtained by adding to the flat target space
action the integrated vertex operator for supergravity massless states and covariantizing
with respect to the ten-dimensional $N=2$ super reparamatrization.
Define by $Z^M = (X^m,\theta^{\a},\bar{\theta}^{\a})$ coordinates on $R^{10|32}$ superspace.
The pure spinor sigma model action takes the form
\be
S = \int d^2z[\frac{1}{2}(G_{MN}(Z)+B_{MN}(Z)) \p Z^M\bar{\p}Z^N +F^{\a\b}d_{\a}\bar{d}_{\b} + ...] + S_{\lambda} + S_{\bar{\lambda}} \ .
\ee
$G_{MN}, B_{MN}, F^{\a\b} ...$ are backgrounds superfields
\be
G_{MN} = g_{MN} + O(\theta),\;\;\;\; B_{MN} = B_{MN} + O(\theta),\;\;\;\; F^{\a\b} = f^{\a\b} + O(\theta) \ .
\ee
Note that $d_{\a}\bar{d}_{\a}$ are independent variables ($p_{\a}, \bar{p}_{\a}$ do not appear explicitly).

The BRST operator reads
\be Q_B = Q + \bar{Q} = \oint dz\,
\lambda^{\alpha}d_{\a} +
\oint
d\bar{z}\,\bar{\lambda}^{{\alpha}}\bar{d}_{{\a}} \ .
\ee
The ten-dimensional supegravity field equations are derived by the
requirement that $Q$ ($\bar{Q}$) is holomorphic (antiholomorphic) and nilpotent
\cite{Berkovits:2001ue}.

\subsection{Curved backgrounds with AdS symmetries} \label{sec:pure-spinor-sigma-model}

Consider pure spinor sigma-models whose target space is the
coset $G/H$, where $G$ is a supergroup with a $\mathbb{Z}_4$
automorphism and the subgroup $H$ is the invariant locus of this
automorphism \cite{Adam:2007ws}. The super Lie algebra $\mathcal{G}$ of $G$ can be
decomposed into the $\mathbb{Z}_4$ automorphism invariant spaces
\be
\mathcal{G} = \mathcal{H}_0 \oplus \mathcal{H}_1 \oplus
\mathcal{H}_2 \oplus \mathcal{H}_3 \ ,
\ee
 where the subscript keeps
track of the $\mathbb{Z}_4$ charge and in particular
$\mathcal{H}_0$ is the algebra of the subgroup $H$. This
decomposition satisfies the algebra $(i=1, \dots, 3)$
\begin{equation}
  [\mathcal{H}_0, \mathcal{H}_0] \subset \mathcal{H}_0 \ , \quad
  [\mathcal{H}_0, \mathcal{H}_i] \subset \mathcal{H}_i \ , \quad
  [\mathcal{H}_i, \mathcal{H}_j] \subset \mathcal{H}_{i + j \
  \mathrm{mod}\ 4} .
\end{equation}
and the only non-vanishing supertraces%
are
\begin{equation}
  \langle \mathcal{H}_i \mathcal{H}_j \rangle \neq 0 \ ,\  i + j = 0 \
  \mathrm{mod}\ 4 \quad (i, j = 0,
  \dots, 3) \ .
\end{equation}
We recall that the supertrace of a supermatrix $M = \left(
\begin{array}{cc}
  A & B \\
  C & D
\end{array}
\right)$ is defined as $\langle M \rangle = \mathrm{Str} M =
\mathrm{tr} A - (-1)^{\mathrm{deg} M} \mathrm{tr} D$, where
$\mathrm{deg} M$ is $0$ for Grassmann even matrices and $1$ for
Grassmann odd ones.

We will denote the bosonic generators in $\mathcal{G}$ by $T_{[a
b]} \in \mathcal{H}_0$, $T_a \in \mathcal{H}_2$, and the fermionic
ones by $T_\alpha \in \mathcal{H}_1$, $T_{\hat \alpha} \in
\mathcal{H}_3$.

It is instructive to consider first the GS action.
The worldsheet fields are the maps $g: \Sigma \to G$ and dividing
by the subgroup $H$ is done by gauging the subgroup $H$ acting
from the right by $g \simeq g h$, $h \in H$. The sigma-model is
further constrained by the requirement that it be invariant under
the global symmetry $g \to \hat g g$, $\hat g \in G$. The
left-invariant current is  defined as
\begin{equation}
  J = g^{-1} d g \ ,
\end{equation}

This current can be decomposed according to the $\mathbb{Z}_4$
grading of the algebra
\be
J = J_0 + J_1 + J_2 + J_3 \ .
\label{deco}
\ee
These currents are manifestly invariant under the global symmetry,
which acts by left multiplication. Under the gauge transformation,
which acts by right multiplication, they transform as
\begin{equation}
  \delta J = d \Lambda + [J, \Lambda] \ , \quad \Lambda \in
  \mathcal{H}_0 \ .
\end{equation}

Using the above properties of the algebra $\mathcal{G}$ and the
requirement of gauge invariance leads to the GS action
\begin{equation} \label{eq:generic-GS-coset-action}
  S_\mathrm{GS} = \frac{1}{4} \int \langle J_2 \wedge *J_2 + J_1
  \wedge J_3 \rangle = \frac{1}{4} \int d^2 \sigma \langle \sqrt{h}
  h^{m n} J_{2 m} J_{2 n} + \epsilon^{m n} J_{1 m} J_{3 n} \rangle \ ,
  \label{eq:coset-GS-action}
\end{equation}
where $m, n = 1, 2$ are worldsheet indices.  A $J_0 \wedge *J_0$
term does not appear because of gauge invariance, while the term
$J_1 \wedge *J_3$ breaks $\kappa$-symmetry and therefore cannot be
included in the GS action. The first and second terms in the
action are the kinetic and Wess-Zumino terms, respectively. The
coefficient of the Wess-Zumino term is determined using
$\kappa$-symmetry.
For a particular
choice of the supergroup, this GS action reproduces the GS action on $AdS_5 \times S^5$ \cite{Metsaev:1998it}.

Let us turn now to the pure spinor sigma-model.
The worldsheet action in the pure spinor formulation of the
superstring consists of a matter and a ghost sector. The
worldsheet metric is in the conformal gauge and there are no
reparameterization ghosts. The matter fields are written in terms
of the left-invariant currents $J = g^{-1}
\partial g$, $\bar J = g^{-1} \bar \partial g$, where $g: \Sigma
\to G$, and decomposed according to the invariant spaces of the
$\mathbb{Z}_4$ automorphism (\ref{deco}).

 The Lie algebra-valued
pure spinor fields and their conjugate momenta are defined as
\begin{equation}
  \lambda = \lambda^\alpha T_\alpha \ ,\quad
  w = w_\alpha \eta^{\alpha \hat \alpha} T_{\hat \alpha} \ , \quad
  \bar \lambda = \bar \lambda^{\hat \alpha} T_{\hat \alpha} \ , \quad
  \bar w = \bar w_{\hat \alpha} \eta^{\alpha \hat \alpha} T_\alpha \
  ,\label{purepara}
\end{equation}
where we decomposed the fermionic generators $T$ of the super Lie
algebra $\mathcal{G}$ according to their $\mathbb{Z}_4$ gradings
$T_\alpha \in \mathcal{H}_1$ and $T_{\hat \alpha} \in \mathcal{H}_3$
and used the inverse of the Cartan metric $\eta^{\alpha \hat
\alpha}$. The spinor indices here are just a reminder, the unhatted
ones refer to left moving quantities, the hatted ones to right moving
ones.  The
pure spinor currents are defined by
\begin{equation}
  N = - \{ w, \lambda \} \ , \quad \bar N = - \{ \bar w, \bar \lambda
  \} \ .
\end{equation}
which generate in the pure spinor variables the Lorentz
transformations that correspond to left-multiplication by elements
of $H$. $N, \bar N \in \mathcal{H}_0$ so they indeed act on the
tangent-space indices $\alpha$ and $\hat \alpha$ of the pure
spinor variables as the Lorentz transformation. The pure spinor
constraint reads
 \be
 \{\l,\l\}=0,\qquad \{\bar\l,\bar\l\}=0 \ .
 \ee

The sigma-model should be invariant under the global
transformation $\delta g = \Sigma g$, $\Sigma \in \mathcal{G}$.
$J$ and $\bar J$ are invariant under this global symmetry. The
sigma-model should also be invariant under the gauge
transformation
\begin{eqnarray}
  \delta_\Lambda J & = & \partial \Lambda + [J, \Lambda] \ ,\quad
  \delta_\Lambda \bar J = \bar \partial \Lambda + [\bar J, \Lambda]
  \,\quad
  \delta_\Lambda \lambda = [\lambda, \Lambda] \ , \quad
  \delta_\Lambda w = [w, \Lambda] \ , \nonumber\\
  \delta_\Lambda \bar \lambda & = & [\bar \lambda, \Lambda] \ ,
  \quad
  \delta_\Lambda \bar w = [\bar w, \Lambda] \ ,
\end{eqnarray}
where $\Lambda \in \mathcal{H}_0$.

The BRST operator reads
\begin{equation}
  Q = \oint \langle dz \lambda J_3 + d\bar z \bar \lambda \bar J_1
  \rangle \ ,
\end{equation}
 The gauge invariant
BRST-invariant sigma-model takes the form
\begin{equation} \label{eq:AdS-ps-action}
  S = \int d^2 z \left\langle \frac{1}{2} J_2 \bar J_2 + \frac{1}{4} J_1
  \bar J_3 + \frac{3}{4} J_3 \bar J_1 + w \bar\partial  \lambda + \bar w
  \partial \bar \lambda + N \bar J_0 + \bar N J_0 -N \bar N
  \right\rangle \ .
\end{equation}
This result holds for all dimensions and matches the critical $AdS_5
\times S^5$ pure spinor superstring action \cite{Berkovits:2004xu}.

Let us briefly comment on the relation between the pure spinor
action (\ref{eq:AdS-ps-action}) and the GS action
(\ref{eq:generic-GS-coset-action}). The latter, when written in
conformal gauge, reads
\begin{equation}
  S_\mathrm{GS} = \int d^2 z \langle \frac{1}{2} J_2 \bar J_2 +
  \frac{1}{4} J_1 \bar J_3 - \frac{1}{4} J_3 \bar J_1 \rangle \ .
\end{equation}
To this one has to add a term which breaks $\kappa$-symmetry and
adds kinetic terms for the target-space fermions and coupling to
the RR-flux $F^{\a\b}$
\begin{equation}
  S_\kappa = \int d^2 z (d_\alpha \bar J_1^\alpha + \bar d_{\alpha}
  J_3^{\alpha} + F^{\alpha\b} d_\alpha \bar d_{
  \b}) = \int d^2 z \langle d \bar J_1 - \bar d J_3 + d \bar d
  \rangle \ ,
\end{equation}
where, in curved backgrounds, the $d$'s are the conjugate
variables to the superspace coordinates $\theta$'s. After
integrating out $d$ and $\bar d$ we get the complete matter part
\begin{equation}
  S_\mathrm{GS} + S_\kappa = \int d^2 z \langle \frac{1}{2} J_2 \bar
  J_2 + \frac{1}{4} J_1 \bar J_3 + \frac{3}{4} J_3 \bar J_1 \rangle \ .
\end{equation}
This has to be supplemented with kinetic terms for
the pure spinors and their coupling to the background
 \begin{equation}
 S_{gh}=\int d^2 z \left\langle  w \bar\partial  \lambda + \bar w
  \partial \bar \lambda + N \bar J_0 + \bar N J_0 -N \bar N
  \right\rangle
\end{equation}
in order to obtain the full superstring sigma-model
(\ref{eq:AdS-ps-action}) with action $S=S_{GS}+S_\kappa+S_{gh}$.

\subsection{Quantum aspects}

The pure spinor action  (\ref{eq:AdS-ps-action}) is classically gauge invariant
under the right multiplication $g\to gh$, where $h\in H$. In the following we will study
the quantum properties.
We will show that we can always add a local counterterm such that the
quantum effective action remains gauge invariant at the quantum
level. Quantum gauge invariance will then be used to prove
BRST invariance.

\subsubsection{Quantum gauge invariance}

An anomaly in the $H$ gauge invariance would show up as a
nonvanishing gauge variation of the effective action
$\delta_\Lambda S_{eff}$ in the form of a local operator. Since
there is no anomaly in the global $H$ invariance, the variation
must vanish when the gauge parameter is constant and, moreover, it
must have grading zero. Looking at the list of our worldsheet
operators, we find that the most general form of the variation is
 \bea
 \delta S_{eff}&=&\int d^2z\langle c_1 N\bar\partial \Lambda+\bar c_1\bar
 N\partial \Lambda+2c_2J_0\bar\partial\Lambda+2\bar c_2\bar
 J_0\partial\Lambda\rangle,\label{lambdas}
 \eea
where $\Lambda=T_{[ab]}\Lambda^{[ab]}(z,\bar z)$ is the local
gauge parameter and $(c_1,\bar c_1,c_2,\bar c_2)$ are arbitrary
coefficients. By adding the counterterm
 \bea
 S_c&=&-\int d^2z\langle c_1 N\bar J_0+\bar c_1\bar NJ_0+(c_2+\bar
 c_2)J_0\bar J_0\rangle,
 \eea
we find that the total variation becomes
 \bea
 \delta_\Lambda (S_{eff}+S_c)&=&(c_2-\bar c_2)\int d^2z\langle
 J_0\bar\partial\Lambda-\bar J_0\partial\Lambda\rangle.
 \eea
On the other hand, the consistency condition on the gauge anomaly
requires that
 \bea
( \delta_\Lambda\delta_{\Lambda'}-\delta_{\Lambda'}\delta_\Lambda)
S_{eff}=\delta_{[\Lambda,\Lambda']}S_{eff},
 \eea
which fixes the coefficients $c_2=\bar c_2$.
Therefore the action is gauge invariant quantum mechanically.

\subsubsection{Quantum BRST invariance}
\label{quantumbrst}

First, we
will show that the classical BRST charge is nilpotent. We will
then prove that the effective action can be made classically BRST
invariant by adding a local counterterm, using triviality of a
classical cohomology class. Then we will prove that order by order
in perturbation theory no anomaly in the BRST invariance can
appear.

As we have shown in the previous section, the action
(\ref{eq:AdS-ps-action}) in the pure spinor formalism is
classically BRST invariant. Also, the pure spinor BRST charge is classically nilpotent
on the pure spinor constraint, up to gauge invariance and the
ghost equations of motion.

Consider now the quantum effective action $S_{eff}$. After the
addition of a suitable counterterm, it is gauge invariant to all
orders. Moreover, the classical BRST transformations
commute with the gauge transformations, since the
BRST charge is gauge invariant. Therefore, the anomaly in the
variation of the effective action, which is a local operator, must
be a gauge invariant integrated vertex operator of ghost number
one
 \bea
  \delta_{BRST}S_{eff}&=&\int d^2z\langle {\Omega}_{z\bar z}^{(1)}\rangle.
  \eea
One can show that the cohomology of such
operators is empty, namely that we can add a local counterterm to
cancel the BRST variation of the action \cite{Berkovits:2004xu,Adam:2007ws}. A crucial step in the
proof is that the symmetric bispinor, constructed with the product
of two pure spinors, is proportional to the middle dimensional
form (\ref{cartan}).

Since there are no conserved currents of ghost number two in the
cohomology, that could deform $Q^2$, the quantum modifications to
the BRST charge can be chosen such that its nilpotence is
preserved. In this case, we can set the anti-fields to zero and
use algebraic methods to extend the BRST invariance of the
effective action by induction to all orders in perturbation
theory. Suppose the effective action is invariant to order
$h^{n-1}$. This means that
 $$\tilde Q S_{eff}=h^n\int d^2z\langle {\Omega}^{(1)}_{z\bar
z}\rangle+{\cal O}(h^{n+1}).
$$
The quantum modified BRST operator $\tilde Q=Q+Q_q$ is still
nilpotent up to the equations of motion and the gauge invariance.
This implies that $Q\int\,d^2z\langle\Omega^{(1)}_{z\bar
z}\rangle=0$. But the cohomology of ghost number one integrated
vertex operators is empty, so $\Omega^{(1)}_{z\bar z}=Q
\Sigma^{(0)}_{z\bar z}$, which implies
 \bea
 \tilde Q\left(S_{eff}- h^n\int\,d^2z\langle\Sigma^{(0)}_{z\bar
 z}\rangle\right)&=& {\cal O}(h^{n+1}).
 \eea
Therefore, order by order in perturbation theory it is possible to
add a counterterm that restores BRST invariance.

\subsubsection{Pure spinor beta-functions}

Consider the computation of the beta-function in the pure
spinor formalism in the background field method \cite{Vallilo:2002mh}. Unlike the
light-cone GS formalism, one works covariantly at all stages.
The contribution to the one-loop effective action coming from the
pure spinor sector
consists of two terms. The first term is
obtained by expanding the ghost action $\frac{1}{\lambda^2}\int
d^2z\,\langle N\bar J_0+\bar N J_0\rangle$ to the second order in the
fluctuations of the gauge current $J_0$. The trilinear couplings
 \bea
\int d^2z\, \langle \tilde N\left([\bar \partial X_2,X_2]+[\bar
\partial X_1,X_3]+[\bar
 \partial X_3,X_1]\right)\\
 +\tilde{\bar N}\left([\partial X_2,X_2]+[\partial X_1,X_3]+[\partial
 X_3,X_1]\right)\rangle,
 \eea
 generate the term $\langle \tilde N\tilde {\bar N}\rangle$ in
 the action
 \bea
 {1\over 8\pi}{\rm log}{\Lambda\over \mu} \tilde N^{[ij]}\tilde{\bar N}^{[kl]}\left(
 4R_{[ij][kl]}(G)-4R_{[ij][kl]}(H)\right).\label{ghostfish}
 \eea

There is a second
contribution to the one-loop effective action in the ghost sector,
coming from the operator ${\cal O}(z,\bar z)=\langle N\bar
N\rangle$, which couples the pure spinor Lorentz currents to the
spacetime Riemann tensor. The marginal part of the OPE of ${\cal
O}$ with itself generates at one-loop the following contribution
to the effective action
 \be
 {1\over4\pi}\int d^2z\int d^2w \langle {\cal O}(z,\bar z){\cal O}(w,\bar
 w)\rangle={1\over2\pi}{\rm log}{\Lambda\over \mu}R_{[ij][kl]}(H)\int
 d^2z \tilde N^{[ij]}\tilde{\bar N}^{[kl]},
 \ee
which cancels the term proportional to $R_{[ij][kl]}(H)$ in
(\ref{ghostfish}). So we are left with the following ghost
contribution to the one-loop effective action in the ghost sector
\bea
 {1\over2\pi}{\rm log}{\Lambda\over \mu} \tilde N^{[ij]}\tilde{\bar N}^{[kl]}
 R_{[ij][kl]}(G) \ ,
 \eea
where the explicit expression of the super Ricci tensor of the
supergroup in terms of the structure constants is explained in the
appendix. When the supergroup $G$ is super Ricci flat, each coupling in
the effective action vanishes by itself, all of them being
separately proportional to the dual Coxeter number of the
supergroup $G$. However, for non-critical superstrings, in
which the dual Coxeter number of $G$ is nonzero the
single terms do not vanish separately and to check that the total contribution vanihses.

\section{Pure Spinor Superstrings in Various Dimensions}

In this section we will consider noncritical pure spinor superstrings \cite{Adam:2006bt}.

\subsection{Noncritical superstrings}

The critical dimension for the superstrings in flat space-time is
$d=10$. In dimensions $d<10$, the Liouville mode is dynamical, i.e. with $g_{ab}=e^{\vp}\delta_{ab}$, the conformal (Liouville) mode
$\vp$ does not decouple and
needs to be quantized as well. These superstrings are sometimes
called noncritical. The Liouville mode can be interpreted as a
dynamically generated dimension. Thus, if we start with
superstring theory in  $d<10$ space-time dimensions, we have
effectively $d+1$ space-time dimensions. The total conformal
anomaly vanishes  for the non-critical superstrings due to the
Liouville background charge. However, while this is a necessary
condition for the consistency of non-critical superstrings, it is
not a sufficient one.

There are various  motivations to study noncritical  strings.
First, non-critical  superstrings can provide alternative to
superstring compactifications. Second, the study of noncritical
superstrings in the context of the gauge/string correspondence may
provide dual descriptions of new gauge theories, and in particular
QCD.
In this context, one would like to study backgrounds with warped type metrics of the form
\be
ds^2 = d\vp^2 + a^2(\vp)dx_idx^i \ .
\ee
The form of the warp factor $a^2(\vp)$ determines the type of the dual
gauge theory, e.g. with  $a(\vp)\sim e^{\vp}$ one has a conformal model, while
with a warp factor vanishing at a point $a(\vp=\vp^*)=0$, one has a confining one.

A complication in the study of  non-critical superstrings in
curved spaces is that, unlike the critical case, there is no
consistent approximation where  supergravity provides a valid
effective description. The reason being that the $d$-dimensional
supergravity low-energy effective action contains a cosmological
constant type term of the form
$$
S \sim \int d^d x \sqrt{G}e^{-2 \Phi}\left({d-10 \over
l_s^2}\right) \ ,
$$
which vanishes only for $d=10$. This implies that the low energy
approximation $E\ll l_s^{-1}$ is not valid when $d \neq 10$, and
the higher order curvature terms of the form $\left(l_s^2{\cal
R}\right)^n$ cannot be discarded. A manifestation of this is that
solutions of the  $d$-dimensional noncritical supergravity
equations have typically curvatures of the order of the string
scale $l_s^2{\cal R}\sim O(1)$ when  $d \neq 10$.
An example is the $AdS_d$ backgound with $N$ units of RR d-form $F_d$ flux, where one has
\be
l_s^2{\cal R} = d-10,\;\;\;\; e^{2\phi} = \frac{1}{N^2},\;\;\;\; l_s^2 F_d^2 \sim N^2 \ .
\ee

The examples that we will consider in the following are two types of backgrounds: the linear dilaton background
and $AdS_{2d}$, for
 $d=1,2$.

\subsection{Pure spinor spaces in various dimensions}
\label{puresection}

As we noted before, the definition of
the pure spinors that Cartan and Chevalley give (\ref{CC}) and (\ref{cartan})
in even dimension
$d=2n$ implies that in
$d=2,4,6$ dimensions the pure spinor is an $SO(d)$ Weyl
spinor.
In some cases one needs  more than just one pure spinor
to construct a consistent string theory, since the pure
spinor spaces are dictated by the realization of the supersymmetry
algebra for the type II superstring \cite{Grassi:2005sb}.

\subsubsection{Two-dimensional superstring}

The left moving sector of Type II superstrings in two dimensions
realizes $\mathcal{N}=(2,0)$ spacetime supersymmetry with $2$ real supercharges
$Q_\a$, both of which are spacetime MW spinors of the same chirality,
which are related by an $SO(2)$ R--symmetry transformation ($\a$ is
not a spinor index in this case, but just enumerates supercharges of
the same chirality). The corresponding superderivatives are denoted by
$D_\a$. The supersymmetry algebra reads
 $$
 \{D_\a,D_\b\}=-\delta_{\a\b}P^+ \ ,
 $$ where $P^\pm$ are the holomorphic (antiholomorphic) spacetime
direction of $AdS_2$. The pure spinors are defined such that
$\l^\a D_\a$ is nilpotent, so that the pure spinor condition in
two dimensions reads \bea
\l^\a\l^\b\delta_{\a\b}=0 \ ,\label{2dpurity} \eea which is solved by
one Weyl spinor.

\subsubsection{Four-dimensional superstring}

In four dimensions, the left moving sector of the type II
superstring realizes ${\cal N}=1$ supersymmetry, which in terms of
the superderivatives $D_A$ in the Dirac form reads
 \bea
 \{D_A,D_B\}&=&-2(C\Gamma^m)P_m \ ,
 \eea
where $C$ is the charge conjugation matrix and $A=1,\ldots,4$.
Requiring nilpotence of $\l^AD_A$ specifies the four-dimensional
pure spinor constraint
 \bea
 \l^A (C\Gamma^m)_{AB} \l^B&=&0 \ .\label{4dpurity}
 \eea

If we use the Weyl notation for the
spinors, under which the pure spinor is represented by a pair of
Weyl and anti-Weyl spinors $(\l^\a,\l^\ad)$, subject to the
constraint
 \bea
 \l^\a\l^\ad=0 \ .
 \label{condition}
 \eea

\subsection{Linear dilaton background}

The $(d+2)$-dimensional linear dilaton background \cite{Kutasov:1990ua}
 \be
\RR^{1,d-1}\times \RR_\vp \times U(1)_x\, ,
 \label{backgr}\ee
has a flat metric in the string frame and a linear dilaton
 $$
 \Phi={Q\over2}\vp \ .
 $$
The effective string coupling $g_s=e^\Phi$ varies as we move along
the $\vp$ direction and when considering scattering processes one
needs to properly regularize the region in which the coupling
diverges. We will only consider the weak coupling region
$\vp=-\infty$, where perturbative string computations are valid.

 The $d+2$ dimensional RNS superstring is
described in the superconformal gauge by $2n+1$ superfields
$X^\mu$, with $\mu=1,\dots,d=2n$, and $X$ and by a Liouville
superfield $\Phi_l$. In components we have $X^\mu=(x^{\mu},
\psi^{\mu})$, $X=(x,\psi_x)$ and $\Phi_l=(\varphi, \psi_l)$, where
the $\psi$'s are Majorana-Weyl fermions.

The $d=2n$ coordinates $x^\mu$ parameterize the even dimensional
flat Minkowski part of the space, while the coordinate $x$ is
compactified on a circle of radius $R=2/Q$, whose precise value is
dictated by the requirement of space-time supersymmetry, as we
will see below. The coordinate $\vp$ parameterizes the linear
dilaton direction with a background charge $Q$. As usual, we need
to add the superdiffeomorphisms ghosts $(\beta, \gamma)$ and
$(b,c)$. The central charge of the system is
 $$
 c=(3/2)(2n+1)_{\{X^\mu,X\}}+(3/2+3Q^2)_{\{\Phi_l\}}+(11)_{\{\b\gamma\}}-(26)_{\{bc\}}
 $$
and the requirement that it vanishes fixes the slope of the
dilaton to $Q(n) = \sqrt{4-n}$. For $n=4$, the background charge
vanishes and we have eight flat coordinates plus $\vp$ and $x$,
getting back to the flat ten-dimensional critical superstring.
When $n\neq 4$ we have noncritical superstrings.

As an example consider
the pure spinor superstring in
the four-dimensional linear dilaton background.
The four-dimensional superstring has $d+1=3$ noncompact directions
$(x^1,x^2,\varphi)$ and the compact $U(1)_x$ direction $x$ with
radius $R=2/Q$, where $Q=\sqrt{3}$ is the Liouville background
charge.
The pure spinor degrees of freedom are given by the  pair of
Weyl and anti-Weyl spinors $(\l^\a,\l^\ad)$ satisfying (\ref{condition}).
The space-time supersymmetry is of half the maximal, i.e. four supercharges
for the closed superstring.

The BRST operator is constructed as
\be
Q = \oint \l^{\a} d_{\a} + \oint \l^\ad d_{\ad}
\ee
in the left sector and a similar one in the right sector.
Note, that only half of the superderivatives in this BRST operator
correspond to true supersymmetries of the linear dilaton background.
This is the reason why the spectrum constructed as the BRST cohomology needs
a projection in order to match the RNS one \cite{Adam:2006bt}.

\subsection{Ramond-Ramond curved backgrounds}

\subsubsection{$AdS_2$} \label{sec:non-critical-AdS_2}

The type IIA non-critical superstring on $AdS_2$ with RR two-form
flux is realized as the supercoset $Osp(2|2)/SO(1,1)\times SO(2)$.
The $Osp(2|2)$ supergroup has four bosonic generators $(\bf
E^\pm,\bf H,\bf \tilde H)$ and four fermionic ones $(\bf Q_\a,\bf
Q_{\ah})$. The index $a=\pm$ denotes the spacetime light-cone
directions. The supercharges are real two-dimensional MW spinors,
the index $\a=1,2$ counts the ones with left spacetime chirality
and the index $\ah=\hat 1,\hat 2$ counts the ones with right
spacetime chirality (note that in the two-dimensional superstring
$\a,\ah$ are not spinor indices but just count the multiplicity of
spinors with the same chirality). To obtain $AdS_2$, we quotient
by $\mathbf{H}$ and $\tilde \HH$, which generate respectively the
$SO(1,1)$ and $SO(2)$ transformations. The $Osp(2|2)$ superalgebra
and structure constants are listed in the  appendix. The left
invariant form $J=G^{-1}dG$ is expanded according to the grading
as
 \bea
 J_0=J^H\HH+J^{\tilde
 H}\tilde\HH,\quad
 J_1=J^\a\QQ_\a,\quad J_2=J^a\EE_a,\quad J_3=J^{\ah}\QQ_{\ah}.
 \eea
and the definition of the supertrace is
 \bea
 \langle
 \EE_a\EE_b\rangle=\delta_{a}^{+}\delta_{b}^{-}+\delta_{a}^{-}\delta_{b}^{+},
 \qquad \langle
 \QQ_\a\QQ_{\ah}\rangle=\delta_{\a\ah},\label{super2}
 \eea
whose details are given in the appendix.

The action of the pure spinor sigma-model is given by
(\ref{eq:AdS-ps-action}), where the pure spinor
$\beta\gamma$-system is defined according to (\ref{purepara}). The
left and right moving pure spinors $\l^\a$ and $\bar \l^\ah$
satisfy the pure spinor constraints (\ref{2dpurity})
 \bea
 \l^\a\delta_{\a\b}\l^\b=0,\qquad \bar \l^\ah\delta_{\ah\bh}\bar\l^\bh=0 \ .
 \eea
Note that the naive central charge counting gives the correct result
\be
c_{tot}=(2)_{\{x\}}+ (-4)_{\{p,\t\}}+(2)_{\{w,\l\}} =0 \ .
\ee

\subsubsection{$AdS_4$}
\label{ads4non}

The non-critical type IIA superstring on $AdS_4$ with RR four-form
flux is realized as a sigma-model on the $Osp(2|4)/SO(1,3)\times
SO(2)$ supercoset. The $Osp(2|4)$ superalgebra and structure constants
are discussed in the appendix. The bosonic generators are the
translations $\mathbf{P}_a$, the $SO(1,3)$ generators
$\mathbf{J}_{ab}$, for $a,b=1,\ldots,4$ and the $SO(2)$ generator
$\mathbf{H}$.  The fermionic generators are the supercharges
$\mathbf{Q}_\a,\mathbf{Q}_\ah$, where $\a,\ah=1,\ldots,4$ are
four-dimensional Majorana spinor indices. We have thus ${\cal N}=2$
supersymmetry in four dimensions. The charge assignment of the
generators with respect to the $\ZZ_4$ automorphism of $Osp(2|4)$ can
be read from the Maurer-Cartan one forms
  \bea
 J_0=J^{ab}{\mathbf{J}_{ab}}+J^{H}\HH,\quad
 J_1=J^\a\QQ_\a,\quad J_2=J^a{\mathbf{P}_a},\quad J_3=J^{\ah}\QQ_{\ah}.
 \eea

The pure spinor sigma-model
is given by (\ref{eq:AdS-ps-action}), where
the pure spinor $\beta\gamma$-system is defined according to
(\ref{purepara}). The left and right moving pure spinors $\l^\a$
and $\bar \l^\ah$ are four-dimensional Dirac spinors, satisfying
the pure spinor constraints (\ref{4dpurity}).
Note that also here the naive central charge counting gives the correct result
\be
c_{tot}= (4)_{\{x\}}+ (-8)_{\{p,\t\}}+(4)_{\{w,\l\}} =0 \ .
\ee

\section{Integrability of Pure Spinor Superstrings}

 A crucial property of sigma-models on supercosets
$G/H$, where the supergroup $G$ has a $\ZZ_4$ automorphism, whose
invariant locus is $H$,
is their classical integrability. In order to exhibit the
integrability of the pure spinor sigma-models, we have to construct an infinite
number of BRST invariant conserved charges.

The
first step in the construction of the charges is to find a
one-parameter family of currents $a(\mu)$ satisfying the flatness
condition
\begin{equation}
da(\mu) + a(\mu) \wedge a(\mu) = 0 \ .\label{laxin}
\end{equation}
One then constructs the Wilson line \be U_{(\mu)}(x,t;y,t) =
\mathrm{P} \exp \left( -\int_{(y,t)}^{(x,t)}a(\mu) \right) \ ,
\label{wilson} \ee and obtains the infinite set of non-local
charges $Q_n$ by expanding \be U_{(\mu)}(\infty,t;-\infty,t) = 1+
\sum_{n=1}^{\infty} \mu^n Q_n \ . \label{nlocal} \ee The conservation
of $Q_n$ is implied by the flatness of $a(\mu)$.

The first two charges $Q_1$ and $Q_2$ generate the Yangian
algebra, which is a symmetry algebra underlying the type II
superstrings propagating on the AdS backgrounds with Ramond-Ramond
fluxes in various dimensions. Moreover, in the pure spinor
formalism one can see that this symmetry holds also at the quantum
sigma-model level.

\subsection{Classical integrability of the pure spinor sigma-model}

In this subsection we will demonstrate the classical integrability
of the action (\ref{eq:AdS-ps-action}).
We has to distinguish between two
cases --- a non-Abelian gauge symmetry $H$ and an Abelian one,
which occurs only in the two-dimensional non-critical
superstrings. We will present the non-Abelian case.
The construction of the flat currents in the case of an Abelian
gauge group is similar.

The equations of motion of the currents $J_i$ are obtained by
considering the variation $\delta g = g X$ under which $\delta J =
\partial X + [J, X]$ and using the $\mathbb{Z}_4$ grading and the
Maurer-Cartan equations, so that we get
\begin{eqnarray}
  \nabla \bar J_3 & = & - [J_1, \bar J_2] - [J_2, \bar J_1] + [N, \bar
  J_3] + [\bar N, J_3] \ ,\\
  \bar \nabla J_3 & = & [N, \bar J_3] + [\bar N, J_3] \ ,\\
  \nabla \bar J_2 & = & - [J_1, \bar J_1] + [N, \bar J_2] + [\bar N,
  J_2] \ ,\\
  \bar \nabla J_2 & = & [J_3, \bar J_3] + [N, \bar J_2] + [\bar N,
  J_2] \ ,\\
  \nabla \bar J_1 & = & [N, \bar J_1] + [\bar N, J_1] \ ,\\
  \bar \nabla J_1 & = & [J_2, \bar J_3] + [J_3, \bar J_2] + [N, \bar
  J_1] + [\bar N, J_1] \ ,
\end{eqnarray}
where $\nabla J = \partial J + [J_0, J]$ and $\bar \nabla J = \bar
\partial J + [\bar J_0, J]$ are the gauge covariant derivatives.
The equations of motion of the pure spinors and the pure spinor
gauge currents are
\begin{eqnarray}
  \bar \nabla \lambda & = & [\bar N, \lambda] \ , \quad
  \nabla \bar \lambda = [N, \bar \lambda] \ ,\\
  \bar \nabla N & = & -[N, \bar N] \ , \quad \nabla \bar N = [N, \bar
  N]  \ . \label{eq:AdS-N-eq}
\end{eqnarray}

We are looking for
a one parameter family of gauge invariant flat currents $a(\mu)$.
The left-invariant current $A = g^{-1} a g$ constructed from the
flat current $a$ satisfies the equation
\begin{equation} \label{eq:AdS-ps-flatness}
  \nabla \bar A - \bar \nabla A + [A, \bar A] +
  \sum_{i=1}^3 \left( [J_i, \bar A] + [A, \bar J_i] \right) = 0 \ .
\end{equation}
$A$ and $\bar A$ can depend on all the currents for which there
are equations of motion so
\begin{equation}
  A = c_2 J_2 + c_1 J_1 + c_3 J_3 + c_N N \ , \quad
  \bar A = \bar c_2 \bar J_2 + \bar c_1 \bar J_1 + \bar c_3 \bar J_3 + \bar c_N \bar N \ .
\end{equation}
By requiring the coefficients of the currents to satisfy
(\ref{eq:AdS-ps-flatness}) one obtains
the solutions
\begin{eqnarray}
  c_2 & = & \mu^{-1} - 1 \ , \quad
  c_1 = \pm \mu^{-1/2} - 1 \ , \quad
  c_3 = \pm \mu^{-3/2} - 1 \ , \quad
  \bar c_2 = \mu - 1 \ , \nonumber \\
  \bar c_1 & = & \pm \mu^{3/2} - 1 \ , \quad
  \bar c_3 = \pm \mu^{1/2} - 1 \ , \quad
  c_N = \mu^{-2} - 1 \ , \quad
  \bar c_N = \mu^2 - 1 \ . \label{eq:AdS-flat-current-params}
\end{eqnarray}
Hence, there exists a one-parameter set of flat currents.

The flat currents are given by the right-invariant versions $a = g
A g^{-1}$ and $\bar{a} = g \bar A g^{-1}$ of the currents $A$ and
$\bar A$ found above. The conserved charges are given by
\begin{equation} \label{eq:ps-conserved-charges}
  U_C = \mathrm{P} \exp \left[ -\int_C \left( dz a + d\bar z
  \bar{a} \right) \right] \ .
\end{equation}
These charges are indeed BRST invariant.

The first two conserved charges can be obtained by expanding $\mu
= 1 + \epsilon$ about $\epsilon = 0$. To simplify the notation we
will consider the right invariant currents
\begin{equation}
  j_i \equiv g J_i g^{-1} \ , \quad
  \bar j_i \equiv g \bar J_i g^{-1} \ , \quad
  n \equiv g N g^{-1} \ , \quad
  \bar n \equiv g \bar N g^{-1} \ .
\end{equation}
Using the expansion in $\epsilon$ one gets
\begin{eqnarray}
  a & = & -\left( \frac{1}{2} j_1 + j_2 + \frac{3}{2} j_3 + 2 n
  \right) \epsilon + \left( \frac{3}{8} j_1 + j_2 + \frac{15}{8} j_3 + 3 n
  \right) \epsilon^2 + O(\epsilon^3) \ , \\
  \bar {a} & = & \left( \frac{3}{2} \bar j_1 + \bar j_2 +
  \frac{1}{2} \bar j_3 + 2 \bar n \right) \epsilon + \left(
  \frac{3}{8} \bar j_1 - \frac{1}{8} \bar j_3 + \bar n \right)
  \epsilon^2 + O(\epsilon^3) \ ,
\end{eqnarray}
whose substitution in (\ref{eq:ps-conserved-charges}) and using
$U_C = 1 + \sum_{n = 1}^\infty \epsilon^n Q_n$ yields
\begin{eqnarray}
  Q_1 & = & \int_C \left[ dz \left( \frac{1}{2} j_1 + j_2 +
  \frac{3}{2} j_3 + 2 n \right) - d\bar z \left( \frac{3}{2} \bar j_1
  + \bar j_2 + \frac{1}{2} \bar j_3 + 2 \bar n \right) \right] \ ,\\
  Q_2 & = & - \int_C \left[ dz \left( \frac{3}{8} j_1 + j_2 +
  \frac{15}{8} j_3 + 3 n \right) + d\bar z \left( \frac{3}{8} \bar j_1
  - \frac{1}{8} \bar j_3 + \bar n \right) \right] + \nonumber \\
  && {} + \int_C \left[ dz \left( \frac{1}{2} j_1 + j_2 + \frac{3}{2} j_3 +
  2 n \right) \bigg|_{(z, \bar z)} - d \bar z \left( \frac{3}{2} \bar
  j_1 + \bar j_2 + \frac{1}{2} \bar j_3 + 2 \bar n \right) \bigg|_{(z,
  \bar z)} \right] \times \nonumber \\
  && {} \times \int_o^{(z, \bar z)} \Bigg[ dz' \left( \frac{1}{2} j_1
  + j_2 + \frac{3}{2} j_3 + 2 n \right) \bigg|_{(z', \bar z')} -
  \nonumber\\
  && {} - d\bar z' \left( \frac{3}{2} \bar j_1 + \bar j_2 +
  \frac{1}{2} \bar j_3 + 2 \bar n \right) \bigg|_{(z', \bar z')}
  \Bigg] \ .
\end{eqnarray}
The first charge $Q_1$ is the local Noether charge. The rest of
the conserved charges, which form the Yangian algebra, can be
obtained by repetitive commutators of $Q_2$.

\subsection{Quantum integrability}

In this subsection we will show that the classically
conserved nonlocal currents
can be made BRST invariant quantum mechanically. In this way we
prove quantum integrability of our type II superstring theories.

Consider the charge that generates the global symmetry with
respect to the supergroup $G$
 \bea
 q\equiv q^AT_A=\int d\sigma j^A T_A,
 \eea
where $j^A$ is the corresponding gauge invariant current. Since
this is a symmetry of the theory, the charge is BRST invariant, so
we find $\e Qj=\partial_\sigma h$, where $h=h^A T_A$ is a certain
operator of ghost number one and weight zero. Classical nilpotence
of the BRST charge implies that $Q h=0$.

Consider the operator $:\{h,h\}:$, where $:\ldots:$ denotes a
BRST invariant normal ordering prescription. If there exists a
ghost number one and weight zero operator $\Omega$, such that
 \bea
 Q\Omega=:\{h,h\}:,\label{omeh}
 \eea
then there is an infinite number of nonlocal charges which are
classically BRST invariant. To prove this, consider the nonlocal
operator
 \bea
 k=:\int_{-\infty}^{+\infty}d\sigma\int_{-\infty}^\sigma
 d\sigma'[j(\sigma),j(\sigma')]:.
 \eea
 Its BRST variation is
 $Qk=2:\int_{-\infty}^{+\infty}d\sigma[j(\sigma),h(\sigma)]:$. On
 the other hand, the BRST transformations are classically
 nilpotent, in fact we find
 $Q(2:[j(\sigma),h(\sigma)]:)=\partial_\sigma:\{h(\sigma),h(\sigma)\}:$.
Since there is an operator $\Omega$ that satisfies
(\ref{omeh}), we have
 \bea
 Q(2:[j,h]:-\partial_\sigma\Omega)=0.
 \eea
In other words, the ghost number one weight one operator
$2:[j,h]:-\partial_\sigma\Omega$ is BRST closed. On the other
hand, the BRST cohomology of ghost number one currents ${\cal
O}^{(1)}_\sigma$ is empty, as we will show below. We conclude that
this operator is BRST exact, namely there exists a $\Sigma^{(0)}$
such that $Q\Sigma^{(0)}=2:[j,h]:-\partial_\sigma\Omega$. But then
the nonlocal charge
 \bea
 \tilde q&=&k-\int_{-\infty}^{+\infty} d\sigma\Sigma,
 \eea
is classically BRST invariant and represent the first nonlocal
charge of the Yangian. By commuting $\tilde q$ with itself one
generates the whole Yangian.

It remains to be shown that the BRST cohomology of ghost number
one currents is trivial. This cohomology, in fact, is equivalent
to the cohomology of ghost number two unintegrated vertex
operators, by the usual descent relation
 \bea
 Q \int d\sigma {\cal
O}^{(1)}_\sigma=0 \Rightarrow Q {\cal
O}^{(1)}_\sigma=\partial_\sigma {\cal O}^{(2)}.
 \eea
At ghost number two we have only two unintegrated vertex operators
that transform in the adjoint of the global supergroup $G$, namely
 \bea
 V_1=g\l\bar\l g^{-1},&\quad& V_2=g\bar \l\l g^{-1}.
 \eea
 Their sum is BRST closed, while their difference is not. Finally,
 we have $ V_1+V_2=Q \Omega^{(1)}$ where
 \bea
 \Omega^{(1)}&=&\half g(\l+\bar\l)g^{-1},
 \eea
 so this classical cohomology class is empty.

Now, suppose that we have a BRST invariant nonlocal charge $q$ at
order $h^{n-1}$ in perturbation theory, namely $\tilde Q
q=h^n\Omega^{(1)}+{\cal O}(h^{n+1})$. $\Omega^{(1)}$ must be a
ghost number one local charge, since any anomaly must be
proportional to a local operator. Nilpotence of the quantum BRST
charge $\tilde Q=Q+Q_q$ implies that $Q\Omega^{(1)}=0$, but the
classical cohomology at ghost number one and weight one is empty,
as shown above, so there exists a current $\Sigma^{(0)}(\sigma)$
such that $Q\int d\sigma\Sigma^{(0)}(\sigma)=\Omega^{(1)}$. As a
result $\tilde Q(q-h^n\int d\sigma\Sigma^{(0)}(\sigma))={\cal
O}(h^{n+1})$. Hence, we have shown that it is possible to modify
the classically BRST invariant charges of
(\ref{eq:AdS-flat-current-params}) such that they remain BRST
invariant at all orders in perturbation theory.

\acknowledgments I would like to thank the organizers and participants of
the RTN Winter School on Strings, Supergravity and Gauge Theories, CERN (2008).
I would also like to thank I. Adam, A. Dekel, A. Grassi, C. Mafra, L. Mazzucato and S. Yankielowicz.

\newpage

\appendix

\section{Superalgebras and Supergroups}
\label{superapp}

In this Appendix we gives some details of the superalgebras
discussed in sections three and four of the lectures.

\subsection{Notations}

The superalgebra
satisfies the following commutation relations:
 \be
 [T_{m},T_{n}]=f^{p}_{mn}T_{p}
 \ee
  \be
 [T_{m},Q_{\a}]=F^{\b}_{m\a}Q_{\b}
 \ee
 \be
 \{Q_{\a},Q_{\b}\}=A^{m}_{\a\b}T_{m}
 \ee
where the $T$'s are the bosonic (Grassman even) generators of a
Lie algebra and the $Q$'s are the fermionic (Grassman odd)
elements. The indices are $m=1,...,d$ and $\a=1,...,D$. The
generators satisfy the following super-Jacobi identities:
 \be
f^{p}_{nr}f^{q}_{mp}+f^{p}_{rm}f^{q}_{np}+f^{p}_{mn}f^{q}_{rp}=0
 \ee
 \be
F^{\g}_{n\a}F^{\d}_{m\g}-F^{\g}_{m\a}F^{\d}_{n\g}-f^{p}_{mn}F^{\d}_{p\a}=0
 \ee
 \be
F^{\d}_{m\g}A^{n}_{\b\d}+F^{\d}_{m\b}A^{n}_{\g\d}-f^{n}_{mp}A^{p}_{\b\g}=0
 \ee
 \be
A^{p}_{\b\g}F^{\d}_{p\a}+A^{p}_{\g\a}F^{\d}_{p\b}+A^{p}_{\a\b}F^{\d}_{p\g}=0
 \ee

Generally we can define a bilinear form
 \be
 <X_{ M },X_{ N }>=X_{ M }X_{ N }-(-1)^{g(X_{ M })g(X_{ N })}X_{ N }X_{ M }=C^{P}_{NM}X_{P}
 \ee
where $X$ can be either $T$ or $Q$ and $P=1,...,d+D$ (say the
first $d$ are $T$'s and the rest $D$ are $Q$'s). $g(X_{M})$ is the
Grassmann grading, $g(T)=0$ and $g(Q)=1$ and $C^{P}_{NM}$ are the
structure constants. The latter satisfy the graded antisymmetry
property
 \be
 C^{P}_{NM}=-(-1)^{g(X_{ M }) g(X_{ N })}C^{P}_{MN}
 \ee

We define the super-metric on the super-algebra as the supertrace
of the generators in the fundamental representation
 \be
 g_{MN}=\Str X_M X_N,\label{supermetric}
 \ee
We can further define raising and lowering rules when the metric
acts on the structure constants
 \be
C_{ M  N P}\equiv g_{ M  S }C^{ S }_{ N P}
 \ee
 \be
C_{ M  N P}=-(-1)^{g(X_{ N })g(X_{P})}C_{ M PN }=-(-1)^{g(X_{ M
})g(X_{ N })}C_{ N  M P}
 \ee
 \be
C_{ M  N P}=-(-1)^{g(X_{ M })g(X_{ N })+g(X_{ N
})g(X_{P})+g(X_{P})g(X_{ M })}C_{P N  M }
 \ee
For a semi-simple super Lie algebra ($|g_{ M  N }|\neq0$ and
$|h_{mn}|\neq0$) we can define a contravariant metric tensor
through the relation
 \be
g_{ M P}g^{PN }=\d^{ N }_{ M }
 \ee

The Killing form is defined as the supertrace of the generators in
the adjoint representation
 \be
K_{ M  N }\equiv (-1)^{g(X_{P})}C^{ S }_{P M }C^{P}_{ S  N
}=(-1)^{g(X_{ M })g(X_{ N })}K_{ N  M }
 \ee
(while on the (sub)Lie-algebra we define the metric
$K_{mn}=f^{p}_{mq}f^{q}_{np}$). Explicitly we have
 \be
K_{mn}=h_{mn}-F^{\b}_{m\a}F^{\a}_{n\b}=K_{nm}
 \ee
 \be
K_{\a\b}=F^{\g}_{m\a}A^{m}_{\b\g}-F^{\g}_{m\b}A^{m}_{\a\g}=-K_{\b\a}
 \ee
 \be
K_{m\a}=K_{\a m}=0
 \ee
The Killing form is proportional to the supermetric up to the
second Casimir $C_2(G)$ of the supergroup, which is also called
the dual Coxeter number
 \be
 K_{MN}=-\,C_2(G)\,g_{MN}.
 \ee

In section 3, we computed the one-loop beta-functions in
the background field method. The sum of one-loop
diagrams with fixed external lines is proportional to the Ricci
tensor $R_{MN}$ of the supergroup. The super Ricci tensor of a
supergroup is defined as
 \bea\label{superi}
 R_{MN}(G)&=&-{1\over 4} f^P_{MQ}f^{Q}_{NP}(-)^{g(X_Q)},
 \eea
and we immediately see that $R_{MN}=-K_{MN}$, in particular, we
can write it as
 \bea
 R_{MN}(G)&=&{C_2(G)\over4}g_{MN}.
 \eea

We considered in section 3 and 4 supergroups $G$
with a $\ZZ_4$ automorphism, whose zero locus we denoted by $H$. The
various RR backgrounds we discussed are realized
as $G/H$ supercosets of this kind. The bosonic submanifold is in
general $AdS_p\times S^q$, where the gauge group
$H=SO(1,p-1)\times SO(q)\times SO(r)$, and the $SO(r)$ factor
corresponds to the non-geometric isometries. The examples we considered are
 $$
 \begin{array}{cccccccccc}
 &    G     & {\rm Algebra} &&p  &  q  &  r &&\#_{susy} &C_2(G)\\
AdS_2 &\, Osp(1|2) & \,B(0|1)& & 2  &  0  &  0&&2&-3\\
AdS_2 &\, Osp(2|2) &  \,C(2)  && 2& 0  &  2&&4&-2\\
AdS_4 &\,Osp(2|4)  & \,C(3) && 4  &  0  &  2&&8&-4\\
AdS_5\times S^5&\,PSU(2,2|4)&\,A(4|4)& &  5  & 5  &  0&& 32&0
\label{tablespaces}
\end{array}
 $$
The superspace notations will be as follows: the letters
$\{M,N,\ldots\}$ refer to elements of the supergroup $G$, while
$\{I,J,\ldots\}$ take values in the gauge group $H$ and finally
$\{A,B,\ldots\}$ refer to elements of the supercoset $G/H$. The
lower case letters denote the bosonic and fermionic components of
the superspace indices, while $\#_{susy}$ is the number of real
spacetime supercharges in the background. Then, we can rewrite the
super Ricci tensor of the supergroup (\ref{superi}) making
explicit the $\ZZ_4$ grading
 \bea
 R_{AB}(G)&=&-{1\over 4} f^C_{AD}f^{D}_{BC}(-)^C-\half f^I_{AD}f^D_{BI}(-)^I \ .
 \eea
In particular, its grading two part is
 \bea
 R_{ab}(G)={1\over4}\left(F^\a_{a\hat\b}F^\bh_{b\a}+F^\ah_{a\a}F^\a_{b\ah}\right)-\half
 f^{i}_{ac}f^{c}_{bi} \ .
 \eea
The Ricci tensor of the
supercoset $G/H$ is given
\be
R_{AB}(G/H)=-{1\over 4}
f^C_{AD}f^{D}_{BC}(-)^{C}-f^I_{AD}f^D_{BI}(-)^{I} \ .
\ee

\subsection{$Osp(2|2)$}

The \textrm{Osp(2$\vert$2)} supergroup corresponds to the
superalgebra $C(2)$. It has a bosonic subgroup $Sp(2)\times SO(2)$
and four real fermionic generators transforming in the ${\bf
4}\oplus{\bf 4}$ of $Sp(2)$. It consists of the super matrices
$\mathbf{M}$ satisfying $\mathbf{M^{st}HM=H}$, where
\begin{displaymath}
\mathbf{H} = \left( \begin{array}{cc|cc}
0 & 1 & 0 & 0 \\
-1 & 0 & 0 & 0 \\
\hline
0 & 0 & 1 & 0 \\
0 & 0 & 0 & 1
\end{array} \right)
\end{displaymath}
The superalgebra is obtained by the commutation relations
$\mathbf{m^{st}H+Hm}=0$, where we parameterize \bea
\begin{array}{ccc} \mathbf{m} &=& \left( \begin{array}{cc|cc}
sl(2) & {} & a & b \\
{} & {} & c & d \\
\hline
e & f & {} & {} \\
g & h & {} & so(2)
\end{array} \right)
\end{array}
&\qquad
\begin{array}{ccc}
\mathbf{m^{st}} &=& \left( \begin{array}{cc|cc}
sl(2)^{t} & {} & e & g \\
{} & {} & f & h \\
\hline
-a & -c & {} & {} \\
-c & -d & {} & so(2)^{t}
\end{array} \right)
\end{array}
 \eea
so that from the condition $\mathbf{m^{st}H+Hm} =0$ we find
\begin{displaymath}
\mathbf{m} = \left( \begin{array}{cc|cc}
sl(2) & {} & a & b \\
{} & {} & c & d \\
\hline
-c & a & {} & {} \\
-d & b & {} & so(2)
\end{array} \right).
\end{displaymath}
The Cartan basis for the $Osp(2|2)$ superalgebra is given by the
following supermatrices. The bosonic generators are
\begin{displaymath}
\bf H= \left(
\begin{array}{cc|cc}
1 & 0 & {} & {} \\
0 & -1 & {} & {} \\
\hline
{} & {} & {} & {} \\
{} & {} & {} & {}
\end{array} \right),
\qquad \bf E^{+}= \left(
\begin{array}{cc|cc}
0 & 1 & {} & {} \\
0 & 0 & {} & {} \\
\hline
{} & {} & {} & {} \\
{} & {} & {} & {}
\end{array} \right),
\qquad \bf E^{-}= \left(
\begin{array}{cc|cc}
0 & 0 & {} & {} \\
1 & 0 & {} & {} \\
\hline
{} & {} & {} & {} \\
{} & {} & {} & {}
\end{array} \right),
\qquad \bf \widetilde{H}= \left(
\begin{array}{cc|cc}
{} & {} & {} & {} \\
{} & {} & {} & {} \\
\hline
{} & {} & 0 & 1 \\
{} & {} & -1 & 0
\end{array} \right),
\end{displaymath}
where $(\bf H,\bf E^\pm)$ are the generators of $sl(2)$ while $\bf
\tilde H$ is the generator of $SO(2)$. The fermionic generators
$(\bf Q_\a,\bf Q_\ah)$ are
\begin{displaymath}
\bf Q_1= \frac{1}{\sqrt{2}}\left(
\begin{array}{cc|cc}
{} & {} & 0 & 1 \\
{} & {} & 0 & 0 \\
\hline
0 & 0 & {} & {} \\
0 & 1 & {} & {}
\end{array} \right),
\qquad \bf Q_{\hat 1}= \frac{1}{\sqrt{2}}\left(
\begin{array}{cc|cc}
{} & {} & 0 & 0 \\
{} & {} & 0 & -1 \\
\hline
0 & 0 & {} & {} \\
1 & 0 & {} & {}
\end{array} \right),
\end{displaymath}
\begin{displaymath}
\bf Q_2= \frac{1}{\sqrt{2}}\left(
\begin{array}{cc|cc}
{} & {} & 1 & 0 \\
{} & {} & 0 & 0 \\
\hline
0 & 1 & {} & {} \\
0 & 0 & {} & {}
\end{array} \right)
\qquad \bf Q_{\hat 2}= \frac{1}{\sqrt{2}}\left(
\begin{array}{cc|cc}
{} & {} & 0 & 0 \\
{} & {} & -1 & 0 \\
\hline
1 & 0 & {} & {} \\
0 & 0 & {} & {}
\end{array} \right)
\end{displaymath}
\\
Finally, the $Osp(2|2)$ superalgebra is given by
 \bea
[\HH,\EE^\pm]=\pm 2\EE^\pm,& [\EE^+,\EE^-]=\HH
,& [\HH,\tilde \HH]=0,\nonumber\\
\,[ \tilde \HH, \EE^\pm ]=0, & [\tilde
\HH,\QQ_\a]=\epsilon_{\a\b}\QQ_\b,& [\tilde
\HH,\QQ_\ah]=\epsilon_{\ah\hat\b}\QQ_{\hat \b},\nn\\
\, [\HH,\QQ_\a]=  \QQ_\a,& [
\HH,\QQ_\ah]=-\QQ_{\hat \a},\label{osp2}\\
\{\QQ_\a,\QQ_\b\}=\half\delta_{\a\b}\EE^+, &
\quad\{\QQ_\ah,\QQ_{\hat
\b}\}=\half\delta_{\ah\hat\b}\EE^-,&\quad\{\QQ_\a,\QQ_\ah\}=\half\delta_{\a\ah}
\HH+\half\epsilon_{\a\ah}\tilde \HH,\nn
\\
\,[\EE^+,\QQ_\a]=[\EE^-,\QQ_\ah]=0,&[\EE^+,\QQ_\ah]=-\delta_{\ah\a}\QQ_\a,&
[\EE^-,\QQ_\a]=-\delta_{\a\ah}\QQ_\ah,\nn
 \eea
We classify the generators according to their $\ZZ_4$ charge
 \be
 \begin{array}{c|cc|cc|cc}
 {\cal H}_0 &{}\,& {\cal H}_1 & {}\,&{\cal H}_2 &{}\,& {\cal H}_3 \\
 &&&&&&\\
 \HH,\tilde \HH &{}& \QQ_\a &{}& \EE^\pm &{}& \QQ_\ah.
  \end{array}
 \ee
In the main text, we realize our $AdS_2$ background by quotienting
with respect to the grading zero subgroup, namely $SO(1,1)\times
SO(2)$. The structure constants are
 \be
 f_{ml}^H=\delta_m^+\delta_l^--\delta_m^-\delta_l^+,\quad
 f_{ml}^{\tilde H}=0,
 \ee
$$f_{Hm}^l=2(\delta_m^+\delta^l_+-\delta_m^-\delta^l_-),\qquad f_{\tilde Hm}^l=0$$
$$F_{\a m}^\ah=\delta_\a^\ah\delta_m^-,\qquad
  F_{\ah m}^\a=\delta_\ah^\a\delta_m^+ $$
$$F_{H\a}^\b=\delta_\a^\b,\qquad
  F_{H\ah}^{\hat \b}=-\delta_\ah^{\hat\b}$$
$$F_{\tilde H\a}^\b=\epsilon_{\a\gamma}\delta^{\gamma\b},\qquad
  F_{\tilde H\ah}^{\hat\b}=\epsilon_{\ah\hat\gamma}\delta^{\hat\gamma\hat\b}$$
$$A_{\a\b}^m=\delta_{\a\b}\delta^{m}_+,\qquad
  A_{\ah\hat\b}^m=-\delta_{\ah\hat\b}\delta^{m}_-,$$
$$A^H_{\a\hat\b}=A^H_{\hat\b\a}=\half\delta_{\a\hat\b},\qquad
  A^{\tilde H}_{\a\hat\b}=A^{\tilde H}_{\hat\b\a}=\half\epsilon_{\a\hat\b},$$

The metric on the supergroup is
 \bea
 g_{mn}&=&\delta_{m}^{+}\delta_{n}^{-}+\delta_{m}^{-}\delta_{n}^{+},\\
 g_{\a\ah}&=&-\eta_{\ah\a}=\delta_{\a\ah},\qquad
 g_{ij}=2\delta_{ij},\nn
 \eea
where $m,n=\pm$, $i,j=H,\tilde H$.

The $OSp(1|2)$ supergroup corresponds to the superalgebra
$B(0|1)$. Its bosonic subgroup is $Sp(2)$ and it has two real
fermionic generators transforming in the ${\bf 2}$ of $Sp(2)$. It
can be easily obtained by the one of the $Osp(2|2)$ supergroup by
simply dropping the generators $\bf \widetilde H$ and
$\QQ_2,\QQ_{\hat 2}$.

\subsection{$Osp(2|4)$}

The supergroup $Osp(2|4)$ corresponds to the superalgebra $C(3)$.
Its bosonic subgroup is $Sp(4)\times SO(2)$ and it has eight real
fermionic generators transforming in the ${\bf 4}\oplus{\bf 4}$ of
$Sp(4)$. We classify the generators according to their $\ZZ_4$
charge
 \be
 \begin{array}{c|cc|cc|cc}
 {\cal H}_0 &{}\,& {\cal H}_1 & {}\,&{\cal H}_2 &{}\,& {\cal H}_3 \\
 &&&&&&\\
 {\bf J_{[ab]}},\tilde \HH &{}& \QQ_\a &{}& {\bf P}_a &{}&
 \QQ_\ah,
  \end{array}
 \ee
where $a=0,\ldots,3$ and $\a,\ah$ are four-dimensional Majorana
spinor indices. In the main text, we realize our $AdS_4$
background by quotienting with respect to the grading zero
subgroup, namely $SO(1,3)\times SO(2)$. The structure constants
are
  \be f_{a b}^{[cd]}=
\half \d_{a}^{[c} \d_b^{d]},\quad
f_{a[bc]}^{d}=-f_{[bc]a}^{d}=\eta_{a[b}\d^{d}_{c]}\ee
$$f_{[ab][cd]}^{[ef]}=\frac{1}{2}\d^{[e}_{[a}\eta_{b][c}\d^{f]}_{d]}
=\frac{1}{2}\left(
 \eta_{bc}\d^{[e}_{a}\d^{f]}_{d}+\eta_{ad}\d^{[e}_{b}\d^{f]}_{c}
-\eta_{ac}\d^{[e}_{b}\d^{f]}_{d}-\eta_{bd}\d^{[e}_{a}\d^{f]}_{c}
\right)$$
$$F^{\bh}_{a\a}=-F^{\bh}_{\a a}=\frac{1}{2}(\g_{a})^{\b}{}_{\a}\d^{\bh}{}_{\b}, \quad
F^{\b}_{a\ah}=-F^{\b}_{\ah
a}=\frac{1}{2}(\g_{a})^{\bh}{}_{\ah}\d^{\b}{}_{\bh}$$
$$F^{\b}_{[ab]\a}=-F^{\b}_{\a[ab]}=\frac{1}{2}(\g_{ab})^{\b}{}_{\a}, \quad
F^{\bh}_{[ab]\ah}=-F^{\bh}_{\ah[ab]}=\frac{1}{2}(\g_{ab})^{\bh}{}_{\ah}$$
$$F_{\tilde H\a}^{\b}=\frac{1}{2}(\g^{5})_{\a}{}^{\b}, \quad
F_{\tilde H\ah}^{\bh}=-\frac{1}{2}(\g^{5})_{\ah}{}^{\bh}$$
$$A^{a}_{\a\b}=(C\g^{a})_{\a\b}, \quad
A^{a}_{\ah\bh}=(C\g^{a})_{\ah\bh}$$
$$A^{\tilde H}_{\a\bh}=-2(\g^{5})_{\a}{}^{\g}(\tilde{C})_{\g\bh}, \quad
A^{\tilde H}_{\ah\b}=2(\g^{5})_{\ah}{}^{\gh}(\tilde{C})_{\gh\b}$$
$$A^{[ab]}_{\a\bh}=-\frac{1}{2}(\tilde{C})_{\a\gh}(\g^{ab})^{\gh}{}_{\bh}, \quad
A^{[ab]}_{\ah\b}=-\frac{1}{2}(\tilde{C})_{\ah\g}(\g^{ab})^{\g}{}_{\b}$$
where $C$ is the charge conjugation matrix of $SO(1,3)$. The
supermetric is given by
 \bea
 g_{ab}&=&\eta_{ab},\qquad g_{\a\bh}=2C_{\a\bh}\\
 g_{[ab][cd]}&=&\eta_{a[c}\eta_{d]b},\qquad g_{\tilde H\tilde H}=2.\nonumber
 \eea

\end{document}